\documentclass[global,twocolumn]{svjour}
\usepackage{graphicx}
\journalname{Applied Physics B}
\begin{document}

\title{Theory of atomic diffraction from evanescent waves}

\author{Carsten Henkel\inst{1} \and 
Hartmut Wallis\inst{2} \and
Nathalie Westbrook\inst{3} \and Chris I. Westbrook\inst{3} 
\and Alain Aspect\inst{3} \and
Klaus Sengstock\inst{4} \and Wolfgang Ertmer\inst{4}}

\institute{Institut f\"ur Physik, 
Universit\"at Potsdam, Am Neuen Palais~10, 
14469 Potsdam, Germany,  
\email{Carsten.Henkel@quantum.physik.uni-potsdam.de},
facsimile: ++49 331 977 1767 
\and
Max-Planck-Institut f\"ur Quantenoptik, Hans-Kopfermann-Stra\ss e~1,
85748 Garching, Germany 
and Sektion Physik, Ludwig-Maximilians-Universit\"at M\"unchen, 
Theresienstra\ss e~37, 
80333 M\"unchen, Germany 
\and
Laboratoire Charles Fabry de l'Institut d'Optique, 
Unit\'e de Recherche Associ\'ee au CNRS, B. P. 147, 
91403 Orsay {\sc cedex}, France 
\and
Institut f\"ur Quantenoptik, Universit\"at Hannover,
Welfengarten~1, 30167 Hannover, Germany (formerly Institut 
f\"ur Angewandte Physik der Universit\"at Bonn, Bonn, Germany)}

\date{received 29 May 1998, accepted 17 june 1999}

\maketitle

\abstract{
We review recent theoretical models and
experiments dealing with the diffraction of neutral atoms
by a reflection grating, formed by a standing evanescent wave.
We analyze diffraction mechanisms proposed for normal and
grazing incidence, point out their scopes and confront the
theory with experiment.\newline
}

\titlerunning{Atomic diffraction theory}
\authorrunning{Henkel et al.}

\section{Introduction}

\begin{figure}[b]
\centerline{\resizebox{\columnwidth}{!}{%
\includegraphics*{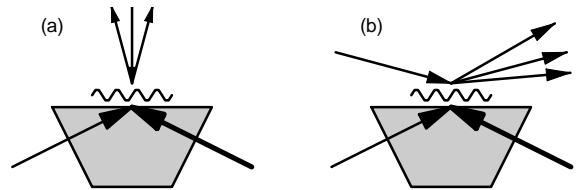}
}}
\caption[fig:scheme]{
Schematical setup of an evanescent wave
reflection grating. (a): normal incidence, (b):
grazing incidence.}
\label{fig:scheme}
\end{figure}

An evanescent wave is the light field formed in vacuum above a
dielectric when a light beam undergoes total internal
reflection at the vacuum--dielectric interface. The evanescent
wave propagates parallel to the interface and decreases
exponentially with distance from the interface, at a scale
sligthly smaller than the optical wavelength. In 1982,
Cook and Hill \cite{Cook82} proposed using this spatially
inhomogeneous light field to construct a repulsive optical
potential able to reflect neutral atoms with normal velocities
(perpendicular to the interface) of several cm/s. 
In 1989, Hajnal and Opat \cite{Opat89}
proposed to combine two counterpropagating evanescent waves,
creating a mirror with a spatially periodic modulation in order to 
realize a reflection grating (cf.\ figure~\ref{fig:scheme}).
At grazing incidence, the diffraction
angles are greatly enhanced similar to X ray diffraction from optical
gratings, and one obtains a beamsplitter useful
for atom optics \cite{Adams94,Wallis95a} and interferometry
\cite{Berman97}.

It took several years of theoretical and experimental work 
\cite{Opat89b,Baldwin94} before efficient diffraction
from a stationary evanescent wave was observed 
\cite{Ertmer94,Brouri96,Landragin97}. 
This is due, on the one hand, to the experimental difficulties:
the diffraction angles are only in the mrad regime because
of the small atomic wavelength compared to the grating period;
the grating has to achieve both reflection and diffraction of
the atomic beam;
at grazing incidence, the coupling between the diffraction orders
is reduced due to the Doppler effect. Diffraction was first achieved
experimentally when all these were overcome: in 1993, in the
experiment at Bonn university (Germany), the atomic beam had been slowed
down \cite{Ertmer94}, while in 1996, the group at Paris-Nord university
(Villetaneuse, France)
tilted the beam with respect to the evanescent wave's propagation direction
\cite{Brouri96}.
At normal incidence, the ENS group (Paris, France) reported 
diffraction in the time domain (1994) \cite{Dalibard95a}, and
in 1996, the group at the Orsay Institut d'Optique
(France) observed that the evanescent wave yields
a nonspecular (diffuse) reflection \cite{Landragin96b}, 
washing out the diffraction peaks, unless in 1997 dielectric surfaces 
with high surface quality were used \cite{Landragin97}.

On the other hand, the theory of the atomic reflection grating 
encountered a number of difficulties: for instance,
the atomic motion is very different from a transmission grating 
because the atoms are also reflected from the grating, 
excluding the elimination of one direction of motion in the
Raman--Nath approximation. 
In 1992, the group at Bonn university solved the Schr\"odinger equation
for a two-level atom in a coupled-wave formalism \cite{Wallis93}
and showed that atomic diffraction is possible 
if the Doppler shift is comparable to both the optical potential 
and the detuning. The proposed diffraction process involves 
a delicate interplay of adiabatically followed dressed states and
nonadiabatic transitions between them. This approach has been used
to interpret numerical calculations \cite{Smith93,Smith94,Savage95}
and allowed a quantitative understanding of experiments 
on Doppleron resonances \cite{Baldwin94}.

For the case of normal incidence and large detunings, the 
group at Institut d'Optique (Orsay, France) developed in 1994 
an alternative theory based on scalar diffraction, 
similar to light diffraction from thin phase objects 
\cite{Henkel94b,Cohen94}. This approach
showed that the atomic wave is very sensitive to spatial modulations
of the evanescent wave because of its small wavelength, in agreement
with the Orsay experiments at normal incidence 
\cite{Landragin97,Landragin96b}. 
At grazing incidence however, the same theory predicts vanishing 
diffraction, in contrast to experiment and the two-level theory
\cite{Wallis93}.

In 1995 the Canberra theory group realized \cite{Savage95} that 
a numerical integration of the coupled wave equations
for a two-level atom at grazing incidence
does not reproduce the Bonn experiment \cite{Ertmer94},
indicating that Doppleron resonances do not provide the relevant
diffraction mechanism.
Since 1996, Gordon and Savage \cite{Savage96}, Henkel et al.\ \cite{Henkel97b}
and Deutschmann \cite{DeutschmannT} pointed out that 
the observed diffraction at grazing incidence 
can only be explained in terms of Raman transitions 
between Zeeman sublevels of the atomic ground state.
A similar mechanism had been studied in 1993 by the Bonn group
\cite{Ertmer93} in view of building a grazing incidence
beamsplitter from a running evanescent wave and a static magnetic field.

The present paper aims to develop a unified picture of these
different theoretical approaches: we review the diffraction mechanisms
mentioned above and compare them to each other and to the diffraction
experiments reported so far \cite{Ertmer94,Brouri96,Landragin97}.
We start in Sec.\ref{s:scalar}
from the most simple description in terms of a `one-level'
optical potential and discuss the results of a perturbative
calculation for the diffraction pattern. For normal incidence
diffraction, this theory has been extended using semiclassical techniques
to compute higher-order diffraction peaks. We then pay particular attention
to the case of grazing incidence. In Sec.\ref{s:two}, the two-level
model of Deutschmann et al.\ is reviewed. We compare it 
to the one-level model and give an interpretation why these models
make different predictions for the diffraction probabilities.
We then apply the two-level model to the diffraction experiments 
and give an analytical estimate of the Doppleron coupling, 
showing it to be too small to account for the observed diffraction patterns. 
On the other hand, the situation is quite different for
Raman couplings, as we discuss finally in Sec.\ref{s:multi}. 
A simple condition for the optimum diffraction efficiency is derived 
and we outline some perspectives for experiments with spin-polarized atoms.

\section{Diffraction of one-level atoms}
\label{s:scalar}

\subsection{Optical potential}

In coherent atom optics, one frequently works in conditions
where the atom is driven non resonantly and at low saturation,
in order to avoid spontaneous emission processes. Under these
circumstances, the theoretical description of the atom--laser-%
interaction simplifies because the excited state may be
eliminated adiabatically. The atomic dynamics is 
governed by a Schr\"odinger equation for the ground state
wavefunction only (`one-level-atom'), where the light field enters via 
the so-called `optical potential' or dipole potential
\begin{equation}
V_{\rm opt}( {\bf r} ) = \frac{ d^2 }{ \hbar \delta }
|{\bf E}( {\bf r} )|^2
\label{eq:opt-pot}
\end{equation}
In this expression, $d$ is the reduced matrix element of the
atomic dipole operator, $\delta = \omega_L - \omega_A$ is the
laser detuning, $\omega_L$ ($\omega_A$) is the laser
(atomic resonance) frequency, respectively, and 
${\bf E}( {\bf r} ) e^{-i\omega_L t} + {\rm c.c.}$
is the laser field. 

To begin with, we consider only a single state in the
atomic ground state. The atomic wavefunction $\psi({\bf r}, t )$
and the Schr\"odinger equation then become scalar quantities 
(without Zeeman sublevel indices)
\begin{equation}
i\hbar \partial_t \psi = 
- \frac{ \hbar^2 }{ 2M } \nabla^2 \psi +
V_{\rm opt}( {\bf r} ) \psi
\label{eq:Schroedinger}
\end{equation} 
The AC-Stark-shift has now acquired the meaning of 
a ponderomotive potential for the center-of-mass motion of 
atoms without internal structure. Such an approach has found widespread
use in theoretical investigations of scalar atom optics,
e.g. for the study of  {\it transmission gratings} (see \cite{Wallis95a}  
for further references).
Although this approach is historically not the first
for the reflection grating
[the first investigations \cite{Opat89,Wallis93} were done for a two-level
atom at arbitrary saturation, see Sec.\ref{s:two}]%
, it gives the most intuitive understanding of the problem. 
Its most straightforward application is 
the description of \emph{`thin'} optical elements, 
i.e. elements which are characterized by the dynamical phase    
the atoms accumulate when passing the optical element. Moreover, 
it allows one to develop far-reaching analytical methods and results that prove
useful in the interpretation of more complicated, even non-scalar models.

\subsection{Kinematics}   

\subsubsection{Reflection grating.}

\begin{figure}[tb]
\centerline{\resizebox{\columnwidth}{!}{%
\includegraphics*{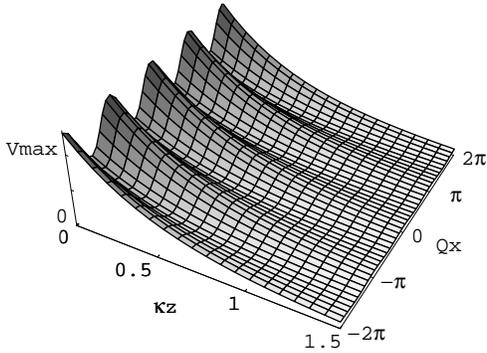}
}}
\caption[fig:EW-pot]{Optical potential~(\ref{eq:EW-pot}) of a partially 
stationary evanescent wave (contrast $\epsilon = 0.3$).}
\label{fig:EW-pot}
\end{figure}

In contrast to a usual transmission grating the atomic 
reflection grating has to achieve two  functions,  
diffraction and reflection, by a suitably shaped optical potential.
A reflective  optical potential with a sharp intensity variation along one 
direction and a periodic intensity  
modulation along the perpendicular direction,
that does this job, can be realized by overlapping
two evanescent waves with wavevectors $Q{\bf e}_x$ and
$-Q{\bf e}_x$ parallel to the dielectric surface
($Q > k \equiv \omega_L / c$). 
Denoting the corresponding field amplitudes by
${\bf E}_\pm( {\bf r} ) = 
{\bf E}_\pm \exp{(\pm i Q x - \kappa z)}$, the optical 
potential~(\ref{eq:opt-pot}) becomes
\begin{equation}
V_{\rm opt}(x,z) =
V_{\max} (1 + \epsilon \cos 2 Q x) e^{-2\kappa z}
\label{eq:EW-pot}
\end{equation}
with
\begin{eqnarray}
V_{\max} & = & \frac{ d^2 }{ \hbar\delta }
( |{\bf E}_+|^2 + |{\bf E}_-|^2 ) 
\\
\epsilon & = & \frac{ 2 {\rm Re}\, {\bf E}_+^*\cdot{\bf E}_- }{
 |{\bf E}_+|^2 + |{\bf E}_-|^2 }
\label{eq:def-contrast}
\end{eqnarray}
The overall height of the potential is given by $V_{\max}$, while
the dimensionless quantity $\epsilon$ determines the \emph{contrast}
of the stationary wave. The potential~(\ref{eq:EW-pot}) is plotted in
Fig.\ref{fig:EW-pot}. The period of the optical potential is equal 
to $\pi/Q = \lambda / (2 n \sin\theta_i)$ and typically a fraction
of the optical wavelength ($n$: index of refraction of the dielectric). 
The same is true for the decay length 
$1/2\kappa = \lambda/[4\pi(n^2\sin^2\theta_i - 1)^{1/2}]$.

Atoms impinging on this optical potential will 
experience a reflection of the $z$-component of their momentum 
if their kinetic energy in this direction does not exceed the potential 
height and they can pick up momentum changes $\pm2Q$ from the periodic 
modulation.

In Eq.(\ref{eq:EW-pot}), we have neglected the (typically gaussian)
intensity profile of the evanescent wave, assuming plane waves
incident inside the dielectric. The present picture also neglects
the van der Waals interaction between the atom and the dielectric
\cite{Landragin96a}.
This interaction is attractive and becomes larger than the optical
potential for distances smaller than roughly $\lambda/4\pi$. The
full potential is well approximated by the optical 
potential~(\ref{eq:EW-pot}) for distances larger than this value.

\subsubsection{Diffraction channels.}

We now introduce the diffraction channels for an
atom incident from infinity onto the stationary evanescent wave.
From energy and momentum conservation for the asymptotic states
(at large distance $z \to \infty$ from the grating),
one finds that the diffracted matter waves are characterized by
the following wavevectors
\begin{eqnarray}
k_{xn} & = & k_{xi} + 2 n Q,
\quad n = 0, \pm1, \pm2, \ldots
\label{eq:def-kxn}
\\
k_{yn} & = & k_{yi}
\label{eq:def-kyn}
\\
k_{zn} & = & \sqrt{k_{zi}^2 - 4 n Q (k_{xi} + n Q) }
\label{eq:def-kzn}
\end{eqnarray}
where the incident wavevector equals $(k_{xi}, k_{yi}, -k_{zi})$.
In the direction parallel to the grating (the $x$-direction),
the atomic wavevector changes by an integer multiple of the
`grating vector' $2Q$. By translation invariance, the $y$ component
is conserved. Finally, the normal (or $z$-) component is obtained
from energy conservation, the optical potential being time-%
independent. These wavevector transfers may also be understood
in the photon picture: in the diffraction process, the atom 
absorbs a photon from one travelling evanescent wave and
emits a (stimulated) photon into the other wave, thereby returning to the
ground state. This can be repeated $n$ times for the $n$th diffraction
order. The momentum
exchange during the process leads to Eqs.(\ref{eq:def-kxn},
\ref{eq:def-kyn}).
Energy is conserved [Eq.(\ref{eq:def-kzn})], corresponding 
to an elastic scattering process,
because the two light waves have the same photon energy (frequency).

From Eqs.(\ref{eq:def-kxn}--\ref{eq:def-kzn}), the
diffraction angles (with respect to the surface normal)
are easily obtained through $\tan\theta_n = k_{xn} / k_{zn}$.
In this article, we focus throughout on two limiting cases: 
normal and grazing incidence. 
At normal incidence and under typical experimental
conditions, the first term in the square root in Eq.(\ref{eq:def-kzn})
dominates the two others: 
the atoms being incident with a velocity much larger than
the recoil velocity $\hbar k / M$,
their incident wavevector $k_{zi}$ is much larger than the optical 
wavevector $k \sim Q$.
 As a consequence, the normal component of the atomic wave vector
changes negligibly, and 
the diffraction angles are small (of order
$\theta_n \simeq 2 n Q/k_{zi}$). The diffracted beams have a
transverse velocity of the order of a few recoil velocities.

At grazing incidence, the second term in the square root,
$-4 n Q k_{xi}$, is comparable to the first, and diffraction leads
to a substantial change of the normal wave vector component. 
The diffraction angles are hence comparable to the angle of
incidence, and the diffraction orders are conveniently separated
in space even for a quite fast atomic beam.
This enormous enhancement of the diffraction angles represents
the main motivation for atomic diffraction at grazing incidence.

Note that the above effect may be interpreted in an alternative
way in the reference frame co-moving with the atomic beam.
In this frame, the frequencies of the two evanescent waves
differ by twice the Doppler shift $\delta_D = Q v_{xi} =
\hbar Q k_{xi}/M$, and diffraction is accompanied by an
energy transfer $2 n \hbar \delta_D$. With this reasoning, one also
finds Eq.(\ref{eq:def-kzn}), up to the `recoil shift' term
$4 n^2 Q^2$ in the root (usually a small correction compared to
the first and second terms).

\subsection{Diffraction intensities: Born approximation}

\subsubsection{Diffraction amplitudes.}

Up to now, we have only specified the kinematics of the
diffraction process. In this subsection, we present a first
approach to calculate the \emph{diffraction pattern},
i.e.\ the fraction of atoms diffracted into the different orders.
To this end, we introduce the asympotic form of
the atomic wavefunction in the region of vanishing potential
(a time-de\-pen\-dent factor $\exp{(-iEt/\hbar)}$ has been separated)
\begin{eqnarray}
&& z \to +\infty: \quad 
\label{eq:psi-asymp}
\\
&&
\psi(x,z) = \exp{i(k_{xi} x - k_{zi} z)} +
\sum_n a_n \exp{i(k_{xn} x + k_{zn} z)}
\nonumber
\end{eqnarray}
The first term represents the incident wave and the sum the
diffracted waves. The coefficients $a_n$ are the \emph{diffraction
amplitudes} that are related to the diffraction probabilities
$\pi_n$ according to
\begin{equation}
\pi_n = |a_n|^2 \frac{{\rm Re}\,k_{zn}}{k_{zi}}
\label{eq:def-pi-n}
\end{equation}
[Diffraction orders with imaginary $k_{zn}$ are evanescent. They
are localized in the optical potential and do not contribute to the
wavefunction at infinity.]

To calculate the diffraction amplitudes $a_n$, one has, in
principle, to solve the stationary Schr\"odinger equation
subject to the asymptotic condition~(\ref{eq:psi-asymp}),
expand the wavefunction in the region $z\to+\infty$ into a
Fourier series with respect to $x$ and read off the expansion 
coefficients. This is a difficult task that may only be 
accomplished numerically, using e.g.\ a coupled-wave expansion
(see Sec.\ref{s:coupled-wave} and 
Refs.\cite{Wallis93,Smith93,Smith94,Savage95}). 
An additional complication arises
from the fact that the atoms are not only diffracted, but also
reflected from the optical potential. This translates into
a second asymptotic condition, namely that
the wavefunction matches an exponentially decreasing solution
inside the potential.
[Corrections to this picture due to tunneling through
the optical potential towards the dielectric are discussed
in \cite{Cote97b}.]
For the same reason, we cannot use Fermi's Golden Rule
to compute the diffraction amplitudes because in this
formula, initial and final states are approximated by
the corresponding asymptotic states (the plane waves of 
Eq.(\ref{eq:psi-asymp})) that are, however, 
a very bad approximation to the true wavefunction deep inside
the potential.

\subsubsection{Distorted-wave Born approximation.}

The way to circumvent this problem and to obtain 
analytical results for the diffraction amplitudes dates back to 
the 1930's \cite{Devonshire36} when Lennard-Jones and Devonshire
calculated the diffraction of a thermal atomic beam from a
periodic (crystalline) surface. The idea is to treat first
the unmodulated potential (that provides the momentum transfer for
the basic reflection process) in an exact manner, and then to
use perturbation theory for the modulated part of the potential.
One obtains a modified Fermi's Golden Rule where wavefunctions
$\psi_{i,f}( z )$ are used that decrease inside the potential and
asymptotically contain an incident and a specularly reflected wave.
The matrix element evaluated in the Golden Rule thus reads
\begin{equation}
A_{fi} = \left\langle \psi_f( z ) \right|
e^{-2 \kappa z} 
\left| \psi_i( z ) \right\rangle
=
\int\limits_{-\infty}^{+\infty}\!{\rm d}z\,
\psi_f^*( z ) e^{-2 \kappa z} \psi_i( z )
\label{eq:matrix-elt}
\end{equation}
This approach to the diffraction problem is called the 
`distorted wave Born approximation' (DWBA) in the following.
It is a perturbative one and of course valid only for a
stationary evanescent wave with low contrast such that 
nonzero diffraction orders are weakly populated.

When the DWBA is applied to the optical potential~(\ref{eq:EW-pot}),
one notes first that the Schr\"odinger equation for the
flat potential (an exponentially increasing potential barrier) 
has an analytical solution \cite{Mott32,Henkel94a}, and second that
the matrix element of the modulated part of the potential may be
evaluated in closed form \cite{Armand79,Henkel94c}. In the 
present notation, one thus obtains the following result for the
diffraction probabilites [neglecting exponentially small corrections
of order $\exp(- \pi k_{zi} / \kappa )$]
\begin{equation}
\pi_{\pm1,\rm DWBA} = \frac{ \epsilon^2 }{ 4 }
\left( \frac{ k_{z,\pm1} + k_{zi} }{ 2\kappa } \right)^2
\beta^2[ ( k_{z,\pm1} - k_{zi} ) / \kappa ]
\label{eq:result-Born}
\end{equation}
where the dimensionless factor $\beta(\xi)$ is defined by
\begin{equation}
\beta( \xi ) = \frac{ \pi \xi / 2 }{ \sinh(\pi \xi / 2) } 
\label{eq:def-beta}
\end{equation}
and reaches its maximum value (unity) for $\xi = 0$.

Note that the DWBA gives the lowest order contribution
to the diffraction pattern. It hence determines only the
intensities of the first-order diffraction peaks ($n =\pm 1$).
Calculations to second order are possible, though much more 
involved because one has to integrate over a continuous spectrum of
intermediate states.

\subsubsection{Physical interpretation.}

\paragraph{Diffraction efficiency.}
We first note that the diffraction process may be quite efficient
even for a small contrast $\epsilon$. Indeed,
if the factor $\beta$ in~(\ref{eq:result-Born})
is close to its maximum value, the diffraction
peaks have a height proportional to the square of the
product $\epsilon k_{zi} / \kappa$ = $2 \pi \epsilon \cos\theta_i/$%
$(\kappa \lambda_{\rm dB})$, and since $\lambda_{\rm dB}$, 
the wavelength of the incident matter wave, is typically 
much smaller than the decay length $1/\kappa$, 
the contrast $\epsilon$ is multiplied by a large number.

To understand this feature, let us consider the potential of
Fig.\ref{fig:EW-pot}. Obviously the contours 
of the potential equalling a given atomic kinetic energy 
are nearly sinusoidal curves in the $xz$ plane.
This suggests to replace the
optical potential of the stationary evanescent wave by
an infinitely high wall with a spatially modulated
position [the `corrugated hard wall' potential used in
atom-surface scattering theory \cite{Steele80,Blinov94}]
\begin{equation}
z( x )= (\epsilon / 2 \kappa) \cos 2 Q x.
\label{eq:}
\end{equation}
The atomic wave acquires
a position-dependent phase shift $\delta \phi( x ) =
2 k_{zi} z( x )$ upon reflection from this barrier. 
The phase of the outgoing wave hence shows a phase
modulation $u \cos2Qx $ with 
a modulation depth $u = \epsilon k_{zi} / \kappa$.
The phase-modulated wave may 
be expanded into sidebands, and the intensity of the
first-order sidebands is given by (for small contrast)
\begin{equation}
\mbox{hard wall:} \quad
\pi_{\pm1} \simeq \frac{ u^2 }{ 4 } =
\frac{ \epsilon^2 }{ 4 }
\left( \frac{ k_{zi} }{ \kappa } \right)^2
\label{eq:hard-wall}
\end{equation}
This coincides with Eq.(\ref{eq:result-Born}) in the regime
$k_{z,\pm1} \approx k_{zi}$ where $\beta \equiv 1$. 
The intensity of the sidebands is comparable to that of the carrier 
when the phase modulation depth approaches unity. 
This means that the amplitude of the hard-wall
corrugation, $\epsilon / 2 \kappa$, must be comparable to the
wavelength of the incident wave. The latter being much 
smaller than the optical wavelength, a low contrast $\epsilon$
is sufficient to deeply modulate the atomic phase.

This picture of course ignores the finite width of the 
optical potential. We show now that this width is related to the
behaviour of the factor $\beta$ in~(\ref{eq:result-Born})
as a function of the \emph{normal wavevector
transfer} $\Delta k_{zn} = k_{zn} - k_{zi}$.

\paragraph{Normal vs.\ grazing incidence.}
The function $\beta(\Delta k_{zn} / \kappa)$ (\ref{eq:def-beta})
rapidly vanishes as soon as the normal wavevector transfer $\Delta k_{zn}$ 
exceeds the evanescent wave decay constant $\kappa$. 
It is close to its maximum value of 1 near normal incidence
where the normal wavevector transfer
$\Delta k_{zn}$ is small and varies quadratically with the
grating vector $Q$ [Eq.(\ref{eq:def-kzn}), recall that
typically $Q \ll k_{zi}$].
On the contrary, at \emph{grazing} incidence,  
$\Delta k_{z,\pm1}$ is much larger than $\kappa$
and $\beta$ decreases exponentially fast:
\begin{equation}
|\Delta k_{z}| \gg \kappa: \quad
\beta(\Delta k_{z}) \simeq
\frac{\pi |\Delta k_z|}{\kappa} 
\exp{\left( - \frac{ \pi |\Delta k_z| }{ 2\kappa } \right)}
\label{eq:beta-cutoff}
\end{equation}
This fact practically induces a cutoff in the diffraction process
for momentum transfers $| \Delta k_{zn} | \gg \kappa$.
In particular, the DWBA predicts a vanishing diffraction efficiency for
one-level atoms at grazing incidence. This feature is in contradiction
with the experiments at grazing incidence \cite{Ertmer94,Brouri96},
and we would like to discuss it in more detail.

Consider the matrix element $A_{fi}$~(\ref{eq:matrix-elt}) that
appears in Fermi's Golden Rule when the DWBA is used.
The $z$-integral is in fact limited to a narrow 
interval of width $1/\kappa$ because in the classically forbidden
region $z \to -\infty$, the wavefunctions $\psi_{i,f}( z )$ vanish
exponentially whereas in the region $z \to +\infty$, the exponential 
potential limits the integral. The integrand in Eq.(\ref{eq:matrix-elt})
is hence an oscillating function due to the interference between
the initial and final waves, with an envelope of width $\sim 1/\kappa$.
Applying an argument familiar from the theory of Fourier transforms, we
may now estimate that $A_{fi}$ is significantly different from zero
only if the difference between the wavevectors $k_{zi}, k_{zf}$
is smaller than the inverse width of the envelope:
\begin{equation}
A_{fi} \ne 0 \quad \Leftrightarrow \quad
|k_{zf} - k_{zi}| \le \kappa
\label{eq:efficient}
\end{equation}
Otherwise stated, the diffraction from a standing evanescent wave 
only provides a `photon momentum' of order $\hbar\kappa$ in the
normal direction.%
\footnote{%
This is not in contradiction with the fact that the reflection
reverses the incident momentum, since reflection is due to the 
flat potential that is treated exactly (to arbitrary order). 
Diffraction, on the contrary, is caused by the modulated part 
of the potential that, to lowest order, has to 
provide the momentum transfer between the diffraction channels.%
}
(For an atomic transmission grating, this was pointed out in 1987 by
Martin et al.\ \cite{Pritchard88a}.)
 The functional form of $A_{fi}$ depends on the envelope of the
integrand in the matrix element. Noting that this envelope
is a smooth function of position, the general
properties of Fourier transforms imply that $A_{fi}$
becomes exponentially small for wavevector
transfers much larger than the limit~(\ref{eq:efficient});
this may also be checked from the exact result~(\ref{eq:result-Born}, 
\ref{eq:beta-cutoff}).
This behaviour is in sharp contrast to the hard-wall potential that 
efficiently provides large wavevector transfers in the normal direction,
as seen in the model of the preceding paragraph.
In their paper \cite{Armand79}, Armand and Manson used the
modulated exponential potential in order to study the
impact of a finite-width interaction potential in atom--surface
scattering, in comparison to the corrugated hard wall
model. They also observed that diffraction decreases if the
potential gets `softer'.

We are hence led to the conclusion that
in the framework of the present (scalar) model, 
no atomic diffraction is predicted at grazing incidence, 
which indicates the
need for an alternative approach in order to interpret the
experimental results \cite{Ertmer94,Brouri96} obtained in this geometry. 
This will be done in Secs.\ref{s:two} and~\ref{s:multi}.

\subsection{Diffraction intensities: semiclassical perturbation method}

In this paragraph, we continue discussing the case of
normal incidence in order to compute higher-order diffraction
peaks. In fact, the sensitivity of diffraction
to weakly modulated potentials calls for a
generalization of the approach discussed so far,
to go beyond the first-order calculation of the DWBA.
The method we present here was developed by Henkel et al.\
\cite{Henkel94b}, elaborating on ideas of Cohen-Tannoudji
\cite{Cohen94}.
It still leads to analytical results for the peak intensities
and provides a simple physical picture for the diffraction process,
substantiating the heuristic approach of the corrugated
hard-wall potential introduced above.
The main difference to the DWBA is that the new approximation
scheme remains valid for much
larger values of the contrast $\epsilon$ 
where the diffraction pattern typically contains a large number of peaks. 
Another important difference to the DWBA is 
to use right from the start semiclassical concepts like trajectories
and phases to compute the diffraction pattern.
This approach is justified by the experimental conditions where
the atomic wavelengths are generally much smaller than the
optical wavelength (the typical scale of the diffraction grating).%
\footnote{%
This remark also shows why the DWBA is useful in its own right; 
namely, it also allows one to study diffraction in the very-low-energy
domain where the atomic wavelength is comparable to
the scales of the diffraction grating. This regime has not yet
been thoroughly explored in experiments, though it should lead
to interesting quantum effects \cite{Henkel96,Cote97}.}

\subsubsection{Outline.}

It is instructive to recall 
the example of light diffraction from an acoustic wave 
with a weak index modulation, as discussed in Ch.~12 of
Born and Wolf \cite{BornWolf}.
After the passage through the acoustic wave,
the light field has been phase-modulated, and the diffraction intensities 
are obtained from a Fourier transform of the field amplitude at the
exit of the interaction zone.
A simple way to calculate the optical phase modulation is to
accumulate the refraction index along straight rays through the acoustic wave.
Note that this `recipe' is less accurate than
the geometrical optics approximation since it discards the deflection
of the rays due to the index gradient. If this deflection
is small, the grating may be called `thin' and for this reason,
the outlined method has been termed the `thin phase grating approximation'
(TPGA). Its results are identical to those obtained in the
Raman--Nath-approximation \cite{Henkel94b,BornWolf}.

At first sight, the TPGA does not seem to apply to the
atomic reflection grating we are interested in, since the
`rays' (classical paths) are substantially distorted 
even for the simple reflection. We now show that the TPGA
may nevertheless be used (see \cite{Henkel94b,Cohen94} for details). 
The basic idea is to compute,
for a weakly modulated diffraction grating, 
the phase shift of the atomic wave to lowest order
in the contrast $\epsilon$.
[This approximation does not necessarily imply only
first-order diffraction because the diffraction pattern will contain
several orders if the resulting phase modulation depth is large.]
In this calculation, we may use some basic concepts of Lagrangian 
classical mechanics. 
Recall that for a given trajectory linking two points, 
the atomic phase is equal to the classical action integral $S_{cl}$,
divided by Planck's constant $\hbar$. 
According to the principle of least action, the classical
trajectory that solves the equations of motion corresponds
to an extremum (actually a minimum) of the action integral.
This property allows us to compute the action $S_{cl}$ 
without explicitly solving the equations of motion:
\emph{we obtain a value for $S_{cl}$ accurate 
to first order even if the action integral is calculated along 
a nearby path}. For the evanescent wave diffraction grating with
a weakly modulated potential,
we thus get the atomic phase modulation from
an integral along the classical path reflected at the \emph{flat} 
(non modulated) optical potential because this path is close to
the `real' path. Using the TPGA in this way, one obtains a simple analytical 
result because both the classical trajectory in the flat potential and the
relevant integral along it are explicitly known \cite{Henkel94a,Levi79,%
Cimmino92}, while the real path is more difficult to calculate.

The perturbative approach just outlined is called
the `thin phase grating approximation' (TPGA) by
analogy to the method used in light diffraction.
The atomic diffraction grating is `thin' if the actual classical 
paths are only slightly perturbed by the modulated potential.
In particular, one has to avoid focal
points and caustics inside the interaction region
(see \cite{Henkel94b} for a detailed discussion).
We note that in neutral atom scattering from crystalline surfaces, 
the TPGA is well known under the name
`trajectory approximation' \cite{Blinov94,Levi79}.
In the field of cold neutron wave optics, similar
approaches have been discussed by Felber et al.\
\cite{Felber96}.

\subsubsection{Discussion.}

\begin{figure}[tb]
\centerline{\resizebox{0.95\columnwidth}{!}{%
\includegraphics*{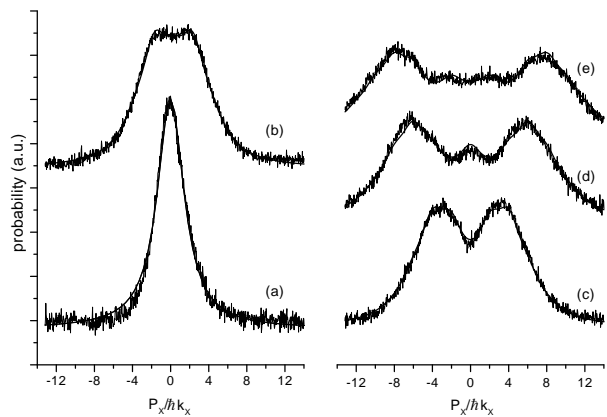}
}}
\caption[fig:bessel]{(courtesy of A. Landragin,
Ref.\cite{Landragin97})
Atomic momentum distributions after normal-incidence
reflection from an evanescent wave with a weak stationary component
(intensity ratio of order $10^{-4}$). 
The amplitude of the standing wave increases from (a) to (e). The solid
line is a theory based upon the thin phase grating approximation:
squared Bessel functions convoluted with the specular reflection 
pattern (a), the modulation index being the only fit parameter.
}
\label{fig:bessel}
\end{figure}

When the thin phase grating approximation is applied to the
atomic diffraction problem, the reflected atomic wave is
phase-modulated with an amplitude
\begin{equation}
u_{\rm TPGA} = \epsilon \frac{ k_{zi} }{ \kappa } 
\beta[ ( 2 Q / \kappa ) \tan\theta_i ]
\label{eq:result-tpg}
\end{equation}
where the cutoff function $\beta$ defined in Eq.(\ref{eq:def-beta})
has been used. 
The diffraction pattern predicted from the TPGA
is thus familiar from phase modulation: the diffraction
peaks have weights given by the squares $J_n^2(u)$
of Bessel functions (see also Fig.\ref{fig:bessel}). 
In the perturbative regime $u \ll 1$, 
the pattern consists of a strong `carrier' (zeroth order peak) plus 
symmetrical sidebands (diffraction orders $n = \pm1$) 
with a weight of order $u^2/4$. 
In the opposite regime $u\gg 1$, sidebands up to order 
$n \approx \pm u$ are populated, with a broad 
maximum at the extreme orders. For a modulation index
$u \approx 2$, the orders $n=\pm1$ are maximized
to a height $\approx 35~\%$ each.

In order to establish the validity of the TPGA, let us
consider the perturbative regime
because there, the DWBA provides us with a `benchmark result'.
At normal incidence, $\theta_i = 0$ and we recover, for a
weak phase modulation $u \ll 1$, the result~%
(\ref{eq:result-Born}) of the DWBA \emph{provided} 
the normal wavevector difference $\Delta k_{z,\pm1}$ 
is small compared to $\kappa$. 
Expanding this difference to second order in $Q$,
we hence find a condition of validity of the thin phase grating
method%
\footnote{%
Note that the TPGA method does
not correctly describe the wavevector transfer normal to the
grating. Roughly speaking, this failure is related to the 
choice of the classical trajectory 
along which the phase shift is computed. In the context
of atomic interferometry, Bord\'e has presented a generalization
of the TPGA that correctly includes the
term quadratic in $Q$ (the recoil term) \cite{Borde97}. Similar
improvements have also been reported for thermal atom scattering
from crystalline surfaces (see \cite{Blinov94} for a review).%
}
\begin{equation}
|\Delta k_{z,\pm1}| \approx \frac{ 2 Q^2 }{ k_{zi} } \ll \kappa,
\label{eq:condition-thin}
\end{equation}
a condition that is typically satisfied in diffraction experiments.
The TPGA hence extends the DWBA result to the regime of large
atomic phase modulation. Its predictions agree 
with diffraction experiments at normal
incidence, as illustrated by Fig.\ref{fig:bessel} taken from
Ref.\cite{Landragin97}.

We conclude this paragraph with an alternative interpretation
for the cutoff of the diffraction efficiency at grazing 
incidence, based on an argument given in 1987 
by Martin et al.\ for a transmission grating \cite{Pritchard87c}.
In the TPGA, one gets the atomic phase modulation 
by accumulating the sinusoidal part of the
potential during the reflection from the flat potential.
At grazing incidence, it happens that the classical path
passes through a large number of standing wave periods during
the interaction time $\tau$.
As a consequence, the standing wave exerts a rapidly oscillating
potential that tends to average out. 
Since this oscillation frequency 
is given by twice the Doppler shift $2 Q v_{xi}$,
we may expect the modulation index to be very small in
the regime $2 Q v_{xi} \tau \gg 1$.
Taking an interaction time $\tau = 1 / (\kappa v_{zi})$)
for the evanescent wave reflection grating,
we recover the argument of the cutoff function
$\beta$ in Eq.(\ref{eq:result-tpg}), and the
asymptotic expansion~(\ref{eq:beta-cutoff}) 
confirms our expectation of a vanishing modulation index.
This cutoff has been observed in experiments performed in the time 
domain \cite{Dalibard95a}, see also \cite{Henkel94c,Felber96},
and very recently in spatial diffraction experiments \cite{Cognet98}.
Balykin et al.\ proposed a method to restore efficient
atomic diffraction at grazing incidence \cite{Balykin98}: 
they reduce the grating vector $2Q$ by using a modulated evanescent wave 
above a structured surface with a period in the $\mu$m range.

\section{Diffraction of two-level atoms}
\label{s:two}

The results of the preceding Section suggest that the one-level model
cannot explain evanescent wave diffraction at grazing incidence 
because the `bandwidth' of this grating is interaction-time limited.
In order to understand why experiments at grazing incidence 
nevertheless showed diffraction,
more complex models are needed and have been introduced, in fact,
prior to the one-level model discussed so far.
These models have in common that they include, to a certain extent,
the atomic multilevel structure. Historically the first approach
to do so is the extensively used two-level atom
\cite{Opat89,Wallis93,Smith93,Smith94,Savage95,Roberts96}
that allows an accurate description of the atom-light
interaction at any degree of saturation. The `next generation'
also includes the magnetic sublevel structure of the atomic
ground state \cite{Savage96,Henkel97b,Ertmer93}.
We discuss in this Section the two-level diffraction theory
developed by Deutsch\-mann et al.\ \cite{Wallis93}.
Its relation to the one-level model is analyzed in some detail,
since it takes a quite different route to explain the
diffraction process.

\subsection{General}

We focus again on the coherent interaction between a two-level atom 
and a monochromatic laser field and assume again that 
spontaneous emission is negligible. The atom is then
described by a two-component wavefunction
$(\psi_g, \psi_e)^T$ whose evolution is governed 
by the following (r.w.a.) Hamiltonian matrix
\begin{equation}
H = - \frac{\hbar^2 }{ 2 M} \nabla^2
- \left(
\begin{array}{cc}
0  & d {E}^*( {\bf r} ) \\
d {E}( {\bf r} ) & \hbar \delta
\end{array}
\right)
\label{eq:2lev-H}
\end{equation}
The notations are identical to Eq.(\ref{eq:opt-pot}),
and we recall the electric field amplitude 
\begin{equation}
{E}( {\bf r} ) = \left(
{E}_+ \exp{i Q x} +  
{E}_- \exp{(- i Q x)}
\right)
e^{- \kappa z}
.
\label{eq:field-amp}
\end{equation}
The laser field couples the ground and excited states,
and therefore new diffraction channels open up where
the atom leaves the interaction region in the excited
state. From momentum conservation, these channels
correspond to an odd number of exchanged photon
momenta along the $x$-axis: $k_{x\nu} = k_{xi} + \nu Q$
with $\nu = 2n+1$.
For the odd diffraction orders, the detuning $\delta$
between the photon energy and the atomic internal energy
has to be taken into account since the atoms end up
in a different internal state (the scattering is
\emph{inelastic}).
Using energy conservation, the normal wavevector component
is obtained as
\begin{eqnarray}
&&\mbox{$\nu$ odd:} \quad 
\label{eq:def-kzn-e}
\\
&&k_{z\nu} = \sqrt{ k_{zi}^2 + 2 M \delta / \hbar 
- \nu Q (2 k_{xi} + \nu Q)}
\nonumber
\end{eqnarray}
For even $\nu$, $k_{z\nu}$ is still given by Eq.(\ref{eq:def-kzn})
with $n = \nu / 2$.

\subsection{Coupled wave analysis}
\label{s:coupled-wave}

To solve the Schr\"odinger equation with the Hamiltonian
(\ref{eq:2lev-H}), it is convenient to use a mixed
position-momentum representation. This takes advantage of
the discrete momentum change along the $x$-direction,
while the dynamics normal to the grating is treated in
the position representation. The wavefunctions are 
thus expanded in Fourier series with respect to the 
$x$-coordinate:
\begin{eqnarray}
\psi_g(x, z) & = &
\sum_{\nu \rm\, even}
\psi_\nu( z ) \exp{i k_{x\nu} x}
\\
\psi_e(x, z) & = &
\sum_{\nu \rm\, odd}
\psi_\nu( z ) \exp{i k_{x\nu} x}
\end{eqnarray}  
In this representation, the coupling term in the
Hamiltonian~(\ref{eq:2lev-H}) becomes an infinite-dimensional 
matrix ${V}_{\nu\mu}$, and the stationary Schr\"odinger
equation takes the following form
\begin{equation}
- \frac{\hbar^2 }{ 2 M} \frac{ d^2 }{ dz^2 } \psi_\nu
+ \sum_\mu {V}_{\nu\mu}( z ) \psi_{\mu}
= \frac{\hbar^2 k_{zi}^2 }{ 2 M } \psi_\nu
\label{eq:Schroed-cw}
\end{equation}
A direct numerical solution of these equations is in principle
possible and has been performed \cite{Opat89,Smith93,Smith94}. 
Recently, also some analytical work has been reported on the
reflection problem \cite{Witte98}.

\subsubsection{Adiabatic potentials.}

\begin{figure}[tb]
\centerline{\resizebox{0.85\columnwidth}{!}{%
\includegraphics*{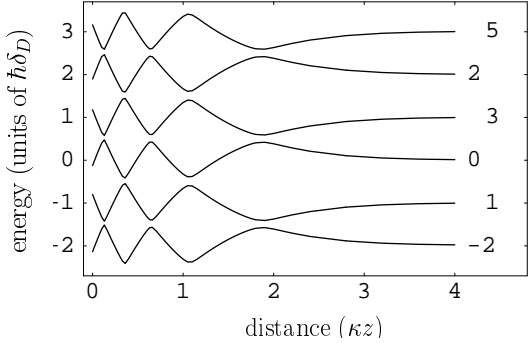}
}}
\caption[fig:pot-ad]{Adiabatic potentials in a stationary
evanescent wave. The potential surfaces are labeled
according to the diffraction channels they asymptotically
connect to. The detuning is $\delta = 2\,\delta_D$.} 
\label{fig:pot-ad}
\end{figure}

In order to simplify the solution and to gain qualitative
understanding, Deutschmann et al.\
diagonalize the coupling matrix ${V}_{\nu\mu}( z )$
in the coupled-wave equations~(\ref{eq:Schroed-cw}). Its
position-dependent eigenvalues $W_\nu( z )$ are called
the `adiabatic' or `dressed state' potentials, where
the latter term draws on the analogy to a two-level atom
in a standing laser wave \cite{Cohen85b}. 
An example of the adiabatic potential surfaces is
given in Fig.\ref{fig:pot-ad}. 

One observes in this figure that far
from the evanescent wave, the potential surfaces are of
two types that are either repulsive or attractive.
For each type, the potentials form an asymptotically
equidistant series with a separation equal to 
twice the Doppler shift.
[Strictly speaking, the levels are not equidistant due to
the `recoil term' quadratic in $Q$, but this is typically
a small correction at grazing incidence.]
For a positive detuning, repulsive potentials 
connect to the ground state channels discussed above for the
`one-level atom', while attractive potentials asymptotically
connect to excited state channels ($\nu$ odd).
Inside the evanescent wave, we observe that repulsive and 
attractive potential surfaces approach each other, although without
crossing. This is because the ground and excited states are coupled
by multiphoton transitions, so-called Doppleron resonances 
\cite{Stenholm77}. 
These `avoided crossings' play a key role
in the diffraction mechanism, as is discussed now.

\subsubsection{Diffraction mechanism.}

Consider a ground state atom that enters
the stationary evanescent wave in the channel $\nu = 0$
and follows the potential $W_0( z )$.
In order to have diffraction, it has to make a transition
to a different potential surface, otherwise it would be
reflected in the $\nu = 0$ channel or hit the surface. 
Such a transition occurs if the atom cannot follow adiabatically the
potential surface $W_0( z )$. This is indeed possible 
because the transformation that diagonalizes 
the coupling matrix $V_{\nu\mu}( z )$ depends on the 
position $z$ and does not commute with the kinetic energy 
operator in the Schr\"o\-din\-ger Eq.(\ref{eq:Schroed-cw}).
The dressed levels thus become  \emph{nonadiabatically
coupled}.
In order to compute the nonadiabatic transitions
between the adiabatic potentials, 
Deutschmann et al.\ exploit the circumstance
that nonadiabatic couplings are only important between levels
that are `close' in energy. As can be seen from Fig.\ref{fig:pot-ad},
this implies that the couplings are spatially localized
around avoided crossings of the adiabatic potentials. In
their vicinity, one may make a two-state approximation and
use a generalization of the Landau-Zener formula  
to compute the wavefunction amplitudes
in the two dressed states after the atom has passed
the avoided crossing (see \cite{Wallis93} for details).%
\footnote{%
The picture presented so far is only accurate in a semiclassical
regime where a generalized Landau-Zener theory for the avoided
crossings applies. It becomes questionable for atoms with a low
kinetic energy because of wave-mechanical effects. In this regime,
Deutschmann et al.\ numerically solved a multi-component
Schr\"odinger equation with appropriate boundary conditions.}

We are thus led to the following picture
for the diffraction process: the incoming atomic wave propagates
through the dressed levels and is split and recombined in the
avoided crossings. This creates a number of partial waves
that propagate either towards the dielectric surface or  
escape into vacuum, after having been reflected from a 
repulsive potential. 
The stationary evanescent wave is hence physically equivalent to
an array of beamsplitters (avoided crossings) and mirrors
(repulsive potentials), 
and the diffraction pattern is determined from the beamsplitters' 
efficiencies and the `optical path lengths' between the splitters 
and mirrors. The atoms traverse this array much like balls 
on a Galton board with the difference that all the different paths
are in principle taken at the same time, their amplitudes
having to be added to obtain the amplitude for diffraction 
into a given order. 

\begin{figure}[tb]
\centerline{\resizebox{0.85\columnwidth}{!}{%
\includegraphics*{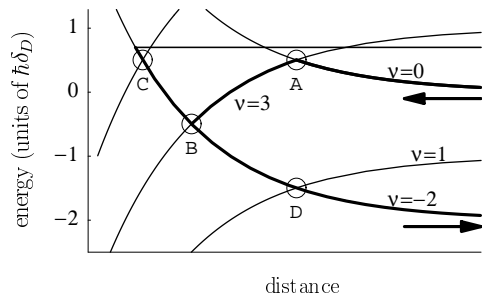}
}}
\caption[fig:4-cross]{Typical path of an atom through the
adiabatic potentials that leads to diffraction into the
$\nu = -2$ ground state channel. The four circles mark 
avoided crossings. The optimized detuning is chosen:
$\delta = 2\,\delta_D$.}
\label{fig:4-cross}
\end{figure}

The performance of the reflection grating is limited by the fact 
that the evanescent wave has to achieve both  diffraction
and reflection of the wavefunction. These tasks are actually
incompatible: for vanishing modulation $\epsilon = 0$
(single running evanescent wave), the
ground state potentials are repulsive, yet there are no 
avoided crossings with attractive potentials. For maximum modulation
$\epsilon = 1$ (a pure standing wave) on the other hand, 
strong avoided crossings
lead to adiabatic potentials widely separated in energy 
and essentially flat; the incoming atoms traverse the
evanescent field and hit the surface. The most efficient operation
of the diffraction grating is obtained for an intermediate contrast.
Another important condition is to achieve an optical potential
comparable to the Doppler shift: this assures that the potential
surfaces for the ground and excited states actually cross. 
The available laser power being limited, one has to reduce 
the Doppler shift $Qv_{xi}$ to meet this condition
\cite{Ertmer94,Brouri96}. At optimized operation, 
the diffraction probabilities are still small, however: 
recall that in an avoided crossing, the atom makes a transition to
an adiabatic potential of excited state character. In order to
end up finally in a ground state diffraction channel,
it must pass several crossings. Deutschmann et al.\ argue 
that at least four crossings are involved, as shown schematically
in Fig.\ref{fig:4-cross}.

\begin{figure}[tb]
\centerline{\resizebox{0.85\columnwidth}{!}{%
\includegraphics*{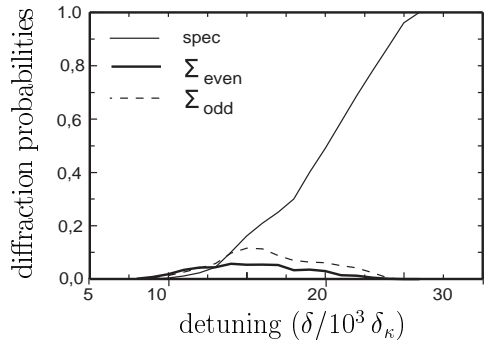}
}}
\caption[fig:ralf]{(courtesy of R. Deutschmann, 
Ref.\cite{Wallis93,DeutschmannT})
Predictions of the two-level model for the populations
as a function of the laser detuning (in units of the `recoil frequency'
$\delta_\kappa \equiv \hbar \kappa^2/2M$). Thin solid curve:
specular ($\nu = 0$) order; thick solid curve: even nonzero
orders; dashed curce: odd orders.
The Doppler shift is taken equal to
$\delta_D = 6.52\times10^{3}\,\delta_\kappa$. 
For a detuning
$\delta = 2 \delta_D$, the even diffraction orders $\nu = \pm 2$ 
are maximized to a total population of about $6~\%$.
The Stueckelberg oscillations are averaged over
by convolution with a finite energy width for the incoming atoms.%
}
\label{fig:ralf}
\end{figure}

Since one of these beamsplitters (marked B) is passed twice, 
first in reflection and second in transmission, 
it is optimized with a $50:50$ splitting ratio. 
Assuming a perfect reflection at the innermost crossing (C)
and a similar splitting ratio for the two outer crossings (A, D),
one is led to a diffraction probability of the order
of $(1/2)^4 \simeq 6~\%$ for the $\nu = -2$ ground state channel.
This value is indeed typical for the results 
obtained in numerical calculations \cite{Wallis93},
as shown in Fig.\ref{fig:ralf}. Diffraction being optimized,
the specular population ($\nu = 0$
channel) is only around $20~\%$ because in a number of crossings,
the atoms are transmitted to the prism surface and lost.
The predicted diffraction probabilities also contain delicate oscillating
features, so-called Stueckelberg oscillations, that are
due to interferences between different paths through the
potential surfaces (not shown in Fig.\ref{fig:ralf}).

\subsection{Comparison to one-level model}

The picture developed by Deutschmann et al.\ seems so
largely different from the one-level atom
that a comparison of the two models is desirable.

First, \emph{why do these models make different predictions
for atomic diffraction at grazing incidence?}
We have seen that in the two-level model, the diffraction
mechanism crucially depends on a resonant coupling to the
excited state playing the role of an intermediate state
for diffraction into a ground state channel. It is 
obvious that the one-level model \emph{cannot reproduce}
this resonant mechanism, since it eliminates the excited state
right from the start, assuming it to be off-resonant. 

While this answers the first question, it leads us to a second one
concerning normal incidence:
\emph{how can we interpret the efficiency of diffraction at 
normal incidence in the adiabatic potential picture?} 
At normal incidence, the Doppler shift is very small, 
and the adiabatic potentials are hence quite close (with a separation
of the order of the recoil energy). A simple estimation
shows that for typical atomic velocities, there
are strong nonadiabatic couplings between the dressed states,
\emph{even outside} avoided crossings \cite{HenkelT}. 
The atomic wavefunction is coupled to other diffraction channels 
all over the evanescent wave, 
and the idea of spatially localized beamsplitters,
that is used in the scheme set up by Deutschmann et al.,
fails in this regime.
By contrast the one-level model does take into account such a
spatially extended coupling. In the thin 
phase grating approximation, e.g., this is done 
by accumulating the phase due to the modulated evanescent wave
potential (that provides the coupling) 
along the path the atom follows during reflection.

Note finally that the preceding argument also allows one to
interpret \emph{why the diffraction efficiency is cut off at grazing
incidence in the scalar model}: with increasing Doppler shift,
the adiabatic potentials for ground-state diffraction channels
become more and more separated, and the probability of nonadiabatic 
transitions decreases exponentially with the level
separation [Kazantsev et al.\ discussed a similar result in 1980
for two-level atoms and a transmission grating \cite{Kazantsev80}].
The atom hence stays adiabatically in its initial repulsive 
dressed level, and no transitions to other (ground state) diffraction
channels take place.

\subsection{Comparison to experiment}

We have to distinguish between two regimes depending on the
value of the detuning compared to the Doppler shift.

The two-level model predicts optimized diffraction for a detuning
$\delta = 2\,\delta_D$ (see Fig.\ref{fig:ralf}). 
Ground and excited states are then
efficiently coupled by a low-order Doppleron
resonance. Recall that in this resonance, 
the atom absorbs a number $l+1$ of photons 
from one running evanescent wave and emits $l$ photons 
into the other, counterpropagating, wave. 
The resonance condition for this process depends on the Doppler shift 
and reads $\delta = (2 l + 1) \delta_D$ in weak fields 
\cite{Stenholm77}; it is shifted towards smaller detunings 
in the presence of light-shift potentials \cite{Baldwin94}.
Optimum diffraction is mediated by a three-photon Doppleron
(the avoided crossing between the channels $\nu = 0, 3$ in 
Fig.\ref{fig:pot-ad}) whose `effective Rabi frequency' 
is comparable to the optical potential.
Once the atom has been excited, however, it is subject 
to an attractive optical potential and likely 
to hit the dielectric surface and be lost from the incident beam.
As discussed above, this is the dominant process for a pure
standing wave where the adiabatic potentials are widely split
at avoided crossings. The two-level model is thus able to
describe the Doppleron resonance experiments 
of the Paris-Nord (Villetaneuse) group \cite{Baudon94} and 
of the Canberra group \cite{Baldwin94} where the evanescent wave's
reflectivity was measured as a function of detuning.

If the detuning is much larger than
the Doppler shift, the two-level model predicts a
vanishing diffraction efficiency (see Fig.\ref{fig:ralf}).
This regime corresponds however to the
diffraction experiments in Bonn \cite{Ertmer94}
and in Paris-Nord (Villetaneuse) \cite{Brouri96}.
There, an atomic beam was split upon reflection 
from a partially stationary evanescent wave. Both groups checked
that the angle of the secondary beam was in accordance with 
the kinematics of a diffraction process. While the Bonn group
observed an efficiency of a few percent, in apparent agreement 
with the two-level theory, the Villetaneuse group reported
a splitting of up to $60:40$ for the $\nu = 0, -2$ channels. 
It is impossible to explain these experiments
with the two-level theory because 
for a detuning $\delta$ much larger than the Doppler shift,
the Doppleron resonances yield 
a negligible diffraction efficiency. This was observed numerically
by Gordon and Savage \cite{Savage96} and by Deutschmann 
\cite{DeutschmannT}, and we would like to give here an analytical
estimate for the Doppleron coupling.

For the experimental parameters of the diffraction experiments
\cite{Ertmer94,Brouri96}, the relevant Dopplerons 
involve a large number of photons ($2 l + 1$ between 10 and 20).
In this regime, eliminating the $2l$ intermediate states,
one finds an `effective Rabi frequency'
of order \cite{HenkelT,Stenholm92b,Wilkens92}
\begin{equation}
\hbar \Omega_{\rm eff} \simeq V_0(z_c) 
\left( \frac{ d E e^{1 - \kappa z_c} }{ \hbar \delta } \right)^{2 l - 1}
\ll  V_0(z_c) 
\label{eq:Rabi-Dopp}
\end{equation}
where $z_c$ is the position of the avoided crossing.
This Rabi frequency is negligible compared to the optical potential 
$V_0(z_c)$ for a detuning
$\delta$ larger than the one-photon Rabi frequency 
$dE e^{-\kappa z_c}/\hbar$.
In the adiabatic potential picture, the avoided crossing becomes a
nearly perfect crossing, and the atom follows its repulsive potential
as if the attractive potential did not exist.
The `beamsplitter' is thus perfectly `transparent',
no splitting and hence no diffraction occur.

\subsection{Summary}

The diffraction of two-level atoms at grazing incidence is possible 
if the atom makes a sequence of transitions that involve, at an
intermediate stage, dressed states of excited character. This
mechanism is effective at low detunings where Doppleron resonances
provide an efficient coupling. For a pure standing evanescent wave,
it leads to a reduced number of reflected atoms because the excited state 
is transmitted down to the dielectric surface. At the large detunings
typical for grazing incidence diffraction experiments, however, Doppleron
resonances become negligible, and  
these experiments call for another model. This has been 
realized by the Bonn group who published, shortly after the
two-level theory, a proposal for an atomic beamsplitter that
involves transitions between ground state Zeeman sublevels
\cite{DeutschmannT,Ertmer93}. 
The Canberra theory group noticed that Zeeman sublevels are
necessary for evanescent wave diffraction 
when they observed that a numerical integration of the coupled-wave 
equations~(\ref{eq:Schroed-cw}) gave negligible diffraction 
for the parameters used in the experiment 
\cite{Savage95,Savage96}.

\section{Model with multiple ground state sublevels}
\label{s:multi}

We now review the diffraction theories that take into account
the Zeeman (magnetic) degeneracy of the atomic ground state.
We first outline the corresponding mechanism
and then summarize the theoretical work done so far.

\subsection{General}

Atomic diffraction occurs when the atom absorbs a photon
from one running evanescent wave and emits one into the
other counterpropagating wave. This `Raman transition'
corresponds to a momentum transfer and leads to the splitting 
of the atomic beam in momentum space. 
At grazing incidence, the running evanescent waves 
that form the diffraction grating acquire a frequency difference 
in the frame of the atomic beam that equals twice the Doppler shift.
If the Raman coupling connects the same ground state sublevel,
it is hence far off resonance and no efficient population
transfer may take place. 
But the Raman transition may also connect different Zeeman sublevels 
and become resonant if the sublevel degeneracy is lifted. 
This happens, e.g., in a static magnetic field (Zeeman effect)
or in a suitably polarized light field. 
The first scenario has been explored 
for a running evanescent wave by Deutschmann et al.\ \cite{Ertmer93},
in view of building an atomic beamsplitter. In the context of
the evanescent wave diffraction grating, the second scenario was studied 
numerically in the Canberra theory group \cite{Savage96}
and analytically in the Orsay group \cite{Henkel97b}. 

\subsection{Physical picture}

For a basic understanding of the diffraction mechanism, let us
consider the limit of low saturation
and a detuning large compared to both the Doppler shift and the
natural linewidth. The excited state manifold may then 
be eliminated adiabatically, and for a ground state of angular momentum $J_g$,
the atomic wavefunction is described by the $2J_g+1$ components 
$\psi_{m}( x, z ), \, m = -J_g, \ldots +J_g$.
It is subject to an optical potential $\hat{V}( z )$
whose matrix elements are of the form
[we suppose that Doppler shift $\delta_D$ 
and Zeeman shift are negligible compared to the detuning $\delta$]
\begin{eqnarray}
&& \left\langle m \right| \hat{V}( z ) \left| m' \right\rangle 
=
\frac{ d^2 }{ \hbar\delta }
\sum_{q,q',m_e}
E_q^*( {\bf r} ) E_{q'}( {\bf r} )
\times
\nonumber\\
&& \quad \times
\left( J_g, m; 1, q \right|\! \left. J_e, m_e \right)
\left( J_e, m_e \right| \!\left. J_g, m'; 1, q' \right)
\label{eq:pot-opt-J}
\end{eqnarray}
where a product of Clebsch--Gordan-coefficients appears on the rhs,
and the electric field is expanded in the usual spherical basis
with coefficients $E_q$, $q = -1, 0, +1$. The optical potential
couples different Zeeman sublevels if the field is 
not in a pure polarization state with respect to this basis. 
The matrix (\ref{eq:pot-opt-J}) then contains both diagonal and nondiagonal
elements. The diagonal elements lead to the light-shift of the 
Zeeman sublevels, that are coupled by the nondiagonal elements.

\begin{figure}[b]
\centerline{\resizebox{0.9\columnwidth}{!}{%
\includegraphics*{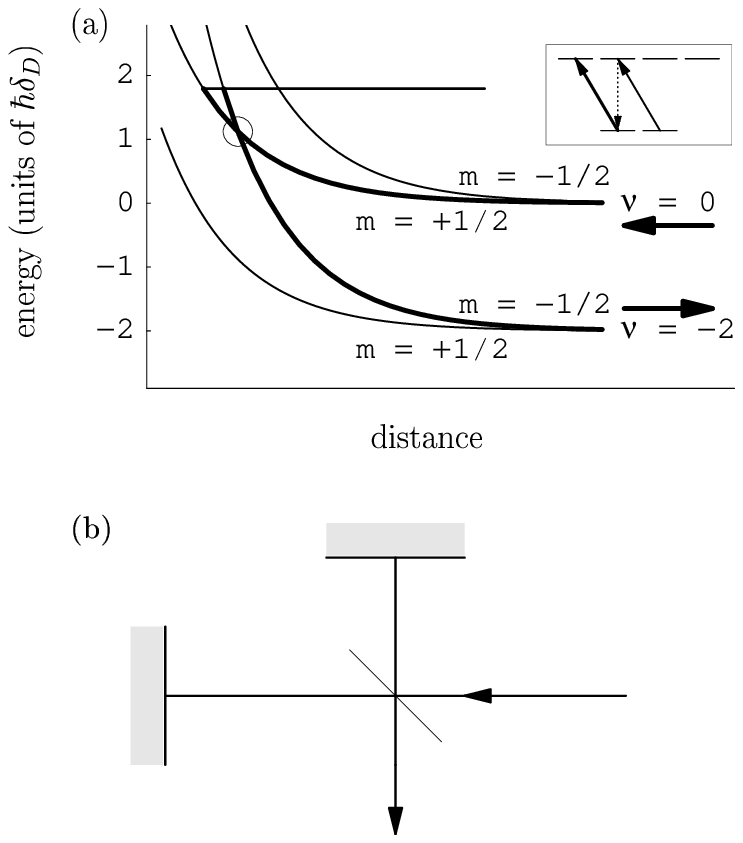}
}}
\caption[fig:pot-ad-J]{(a) Potential energy surfaces for a
$J=1/2 \to J_e = 3/2$ transition in a single (nearly) circularly 
polarized evanescent wave. 
The potentials cross where the difference between 
the light-shifts compensates for the Doppler shift. At this
position, a linearly polarized copropagating wave leads
to a resonant Raman coupling to the $\nu = -2$ diffraction
channel (see inset). The kinetic energy $\frac12 m v_{zi}^2$ of
the incident atoms is indicated by the thin horizontal line.
\newline
(b) Schematic representation of the diffraction process in terms
of a Michelson interferometer. The level crossing of (a) is
represented as a beamsplitter, and the turning points in the
repulsive potentials as mirrors.}
\label{fig:pot-opt-J}
\end{figure}

One may now proceed along similar lines as for the two-level
theory, and compute the adiabatic potentials for the
diffraction channels. These channels are labeled by the
diffraction order $\nu = 0, \pm2, \ldots$ and the magnetic
sublevel $m$. Without a magnetic field and for fixed $\nu$,
the magnetic sublevels are asymptotically degenerate, while different 
diffraction orders $\nu$ are separated by twice the Doppler shift 
(at grazing incidence). An example is shown in 
Fig.\ref{fig:pot-opt-J}(a) for a $J_g = 1/2$ ground state and a
single running evanescent wave with $TM$ polarization
(magnetic field vector perpendicular to the optical plane of incidence). 
To a quite good approximation, this wave is in fact $\sigma^-$ circularly 
polarized provided the laser beam is incident far beyond the critical 
angle. [The quantization axis is parallel to the magnetic field vector.]
The degeneracy of the sublevels is lifted by the light field
because the diagonal elements of the optical potential matrix~%
(\ref{eq:pot-opt-J}) differ in magnitude for the sublevels
$m = \pm 1/2$ [see inset of Fig.\ref{fig:pot-opt-J}(a)]. 
As a consequence, the potentials
for the diffraction orders $\nu = 0, -2$ cross
when the difference between the light shifts 
$\langle -1/2 | \hat{V}( z ) | {-1/2} \rangle $
and $\langle +1/2 | \hat{V}( z ) | {+1/2} \rangle$
is larger than twice the Doppler shift. 
For a pure $TM$ polarization,
this is an exact crossing because there is no Raman coupling 
(the optical potential~(\ref{eq:pot-opt-J})
has no off-diagonal elements).
A coupling can be provided if one adds a second 
evanescent wave with linear ($TE$) polarization:
starting from the sublevel $m = +1/2$,
the atom absorbs a $\sigma^-$ polarized photon
from the strong counterpropagating ($TM$) wave and emits a stimulated
photon with $\pi$ polarization into the weak copropagating ($TE$)
wave. The atom thus ends up in the $m = -1/2$ substate of the
$\nu = -2$ diffraction channel [Fig.\ref{fig:pot-opt-J}(a), inset]. 
In the presence of the second evanescent wave, 
the adiabatic potentials form an avoided crossing at the
position of the circle in Fig.\ref{fig:pot-opt-J}(a).
There, an incoming wavefunction in the $|\nu = 0, m = +1/2\rangle$ channel
is split in two parts that are subsequently reflected 
from their respective repulsive potentials and 
recombined after the second passage at the crossing. Note that
for this particular model, the evanescent wave realizes a `Michelson
interferometer' with a single beamsplitter and two mirrors,
as shown schematically in Fig.\ref{fig:pot-opt-J}(b).

\subsection{Predictions} 

\subsubsection{Numerical calculation.}

In Ref.\cite{Savage96}, the Canberra theory group reported
a numerical solution of the time-dependent Schr\"odinger equation 
for an atom with Zeeman-degenerate ground and excited states 
with $J_g = J_e = 2$. 
They studied a situation close to the experiment \cite{Ertmer94}:
a diffraction grating formed by two evanescent waves 
and an unpolarized atomic beam
(all magnetic sublevels equally populated). 
The population
in the $\nu = -2$ diffraction channel is found of the order of 15~\%
if the optical polarization vectors are tilted (different combinations
of $TE$ and $TM$ polarizations),
in qualitative agreement with the experiment. 
Similar to the magnetic beam splitter~\cite{Ertmer93},
only a single Zeeman substate is involved in the multilevel 
diffraction mechanism. This suggests that a transfer efficiency
close to 100~\% could be possible with spin-polarized atoms.

\subsubsection{Landau-Zener theory.}

In the following, we concentrate on analytical results obtained for
the $J_g=1/2$ ground state discussed above.
The physical picture explained for this model can be translated 
into a simple theory
when the avoided crossing is treated by means of the Landau-Zener
model for nonadiabatic transitions \cite{Zener32}. Assuming that the atom moves
through the crossing with a constant velocity (fixed by energy conservation),
the Landau-Zener formula allows one to compute the probability amplitudes
for the two potentials after the crossing. One thus obtains the
reflection and transmission amplitudes for the beam splitter in
Fig.\ref{fig:pot-opt-J}(b). Calculating the phase shifts between the 
beamsplitter and the turning points in the WKB approximation, we end
up with the populations of the diffraction channels 
$|\nu = 0, m = +1/2\rangle$ and $|\nu = -2, m = -1/2\rangle$.
 
\begin{figure}[tb]
\centerline{\resizebox{0.9\columnwidth}{!}{%
\includegraphics*{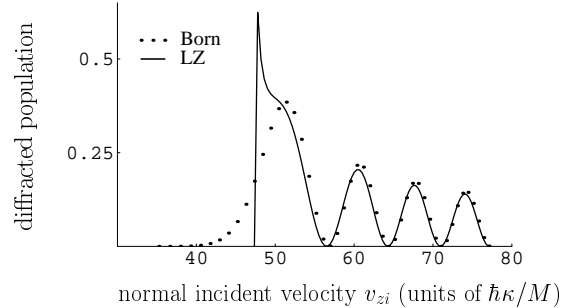}
}}
\caption[fig:Raman-LZ]{(courtesy of C. Henkel, Ref.\cite{Henkel97b})
Predictions of the Landau-Zener model
(solid line) and the distorted-wave Born approximation (points)
for the diffraction of a $J_g=1/2$ atom from an evanescent wave
diffraction grating with a weak polarization gradient. The 
diffracted population is shown as a function of the incident
velocity component $v_{zi}$ (in units of the `recoil velocity'
$\hbar\kappa/M$). The intensity $|{\bf E}_{TE}(z=0)|^2$
of the $TE$ polarized wave is
equal to $6 \times 10^{-4}$ the $TM$ intensity.%
}
\label{fig:Raman-LZ}
\end{figure}

The result of such a calculation is shown in Fig.\ref{fig:Raman-LZ}
(solid line), as a function of the incident
velocity $v_{zi}$ perpendicular to the grating.
The incoming atom is in the substate $m = +1/2$, the
diffracted atom being in $m = -1/2$.
The diffracted population shows a maximum, and this happens 
if the position of the avoided crossing coincides with the
atomic turning point. Since the atom then
spends a long time in a region of resonant Raman coupling, this could
have been expected.%
\footnote{
If the atomic velocity is precisely zero at the crossing, Landau-Zener
predicts in fact that the atom adiabatically follows its initial
potential surface, leading to zero diffraction. A very small deviation
from this condition is sufficient, however, to produce a large
transition probability and hence efficient diffraction.%
}
For larger energies, the atomic wave explores
both potential surfaces [cf.\ Fig.\ref{fig:pot-opt-J}(a)], 
and the diffraction populations show Stueckelberg
oscillations due to the interference between the two paths. The diffracted
population eventually decreases because the atomic velocity becomes 
larger and larger at the avoided crossing, reducing the splitting
efficiency.
The coincidence of the atomic turning point with the avoided crossing
allows us to formulate the following condition for optimum diffraction:
\begin{equation}
\frac{ \langle - 1/2 | \hat{V}( z_c ) | {- 1/2} \rangle 
}{ \langle + 1/2 | \hat{V}( z_c ) | {+ 1/2} \rangle }
=
\frac{ E_{zi} + 2 \hbar \delta_D }{ E_{zi} }
\label{eq:optimum}
\end{equation}
Due to the common exponential variation of the optical potentials,
the lhs is in fact independent of $z_c$ and only depends on the
polarization of the evanescent field. Recalling
that the atom must be reflected from both optical potentials, we
recover the prediction of Deutschmann et al.\ \cite{Wallis93}
that a light-shift comparable to the Doppler shift is necessary
for efficient diffraction.

\subsubsection{Distorted-wave Born approximation.}

Although the Landau-Zener model indicates that diffraction 
is optimized if the atomic turning point coincides with the avoided
crossing, this circumstance invalidates the model itself, because
in the vicinity of the turning point one cannot describe the atom by
a classical particle with a given velocity. We may however use an
alternative approach based on the distorted-wave Born approximation.
The Raman coupling is then treated as a perturbation that induces a
coupling between the wavefunctions in the initial and final potentials.
Note that these wavefunctions have asympotically different kinetic
energies (in the normal direction) due to the Doppler effect.
Analyzing the matrix element in Fermi's Golden Rule in
analogy to the Franck--Condon factors familiar from molecular physics,
we expect diffraction to be most efficient when the classical
turning points of both wavefunctions coincide, since their amplitude
is maximum there.%
\footnote{%
This picture yields still another interpretation of the grazing incidence 
cutoff in the one-level model: 
since the initial and final wavefunctions are subject to the same
light shift, their turning points can only coincide if they have the
same energy, which is impossible due to the Doppler effect. The
Franck--Condon overlap is thus far from its maximum value.%
}
It is easy to see that this condition is in fact identical
to~(\ref{eq:optimum}). Fig.\ref{fig:Raman-LZ} shows
the result of the distorted-wave Born approximation (points). An
optimum diffraction is indeed obtained close to the (unphysical) maximum 
of the Landau-Zener model (solid line), and both models are in
good agreement for incident energies above the optimum.

\subsection{Comparison to experiment}

The numerical calculation~\cite{Savage96} of the Canberra theory
group is in fair agreement with the Bonn experiment~\cite{Ertmer94},
given the number of simplifications in the theory (evanescent wave
of infinite size, omission of van der Waals interactions with the
dielectric surface). For a more detailed comparison, experiments
with spin-polarized atomic beams and well-controlled optical
polarizations would be very useful. The theory indeed predicts
a quite efficient population transfer (100~\% do not seem excluded
in principle), with promising applications for atomic interferometry.

We also note that experiments at large angle incidence 
may be `simulated' in a normal incidence geometry, by introducing
a frequency difference between the two counterpropagating evanescent
waves, in a way similar to early diffraction experiments with
bichromatic standing evanescent waves \cite{Opat89b,Baldwin94}.
Experiments in this direction have been performed in the Orsay group
and confirm the multilevel diffraction theory outlined above
\cite{Cognet98}. 
For a quantitative comparison, however, the van der Waals
interaction and, possibly, losses from spontaneous emission have
to be taken into account. A detailed discussion will be published
elsewhere.

\section{Conclusion}

The diffraction of neutral atoms from a stationary evanescent wave
has remained, since its proposal in 1989, a fascinating challenge,
both experimentally and theoretically. In the last decade, important
breakthroughs have been achieved: detailed theoretical predictions
for both grazing and normal incidence as well as successful
experimental observations. Despite the conceptual simplicity of the setup, 
the diffraction mechanism turned out to be quite subtle. In close contact
with the experimental efforts, theory has now evolved towards a picture
where Raman transitions between magnetic sublevels of the atomic ground state 
play a crucial role. This feature endows atomic diffraction with
a richer structure than, e.g., light diffraction \cite{light},
and it also allows one to construct
tunable and nearly lossless atom-optical beamsplitters, using
suitably polarized light fields far off resonance and/or magnetic fields.
The field is still active and one observes an increased interest in
specifically tailored light fields, using microfabricated surfaces
\cite{Balykin98,Roberts96}, for grazing incidence reflection gratings.
At normal incidence, it has become apparent that atoms moving in an
evanescent wave provide a sensitive probe of van der Waals like 
surface interactions \cite{Landragin96a}. The combination with 
multilevel diffraction mechanisms suggests to probe this interaction
with interferometric resolution and for well-defined Zeeman sublevels.

Another direction of current research may be termed `coherent
atom optics' where one studies the motion of a high-density (and ultimately
Bose--Einstein condensed) atomic gas in laser fields. For instance,
the evanescent wave mirror combined with gravity could be used to build 
an atomic resonator \cite{Wallis92,Wallis96b} where diffraction may
serve as a convenient output coupler. At high densities in this resonator,
one has to take into account the large refraction index of the trapped gas, 
modifying the evanescent wave \cite{Wallis97}. 
A second example are atomic waveguides in hollow fibers: `coating' the walls
with an evanescent wave, efficient guiding has already been
demonstrated \cite{Renn96,Ohtsu96b,Ertmer98}. 
Grazing incidence diffraction taught us that Raman
couplings may mix the ground state sublevels and transfer large
amounts of kinetic energy perpendicular to the wall. This problem
will have to be faced on the route rowards single-mode atomic waveguides.

\paragraph{Acknowledgments.}

\sloppy
\begin{small}
C. H. acknowledges support from the Minist\`ere de la Re\-cherche
et de l'Ensei\-gne\-ment Sup\'e\-rieur (France), from the
Training and Mobility of Researchers Network `Coherent Matter
Wave Interactions' (European Union, contract ERBFMRX CT96-0002),
and from the Deutsche Forschungs\-gemeinschaft (Germany). 

This work would have been impossible without fruitful discussions
with many colleagues among whom we would like to mention
Rosa Brouri,
Laurent Cognet,
Jean-Yves Courtois,
Jean Dalibard,
Ralf Deutsch\-mann,
Gabriel Horvath,
Robin Kaiser,
Arnaud Landragin,
Vincent Lorent,
Klaus M{\o}lmer,
Tilman Pfau, and
Andrew Steane.

\end{small}



\end{document}